\documentclass[twocolumn]{aastex62}
\usepackage{amsmath, amsthm, amssymb, amsfonts, natbib}

\newcommand{\um}{\mu\textnormal{m}}

\graphicspath{{./}{./}}

\received{December 13, 2022}
\revised{December 27, 2022}
\accepted{December 29, 2022}
\submitjournal{Astrophysical Journal Letters}

\shorttitle{JWST NIRSpec Phase Curve For WASP-121\lowercase{b}}
\shortauthors{Mikal-Evans et al.}

\begin{document}

\title{A \textit{JWST} NIRSpec Phase Curve for WASP-121b: Dayside Emission Strongest Eastward of the Substellar Point and Nightside Conditions Conducive to Cloud Formation}

\correspondingauthor{Thomas Mikal-Evans}
\email{tmevans@mpia.de}

\author[0000-0001-5442-1300]{Thomas Mikal-Evans}
\affil{Max Planck Institute for Astronomy, K\"{o}nigstuhl 17, D-69117 Heidelberg, Germany}

\author[0000-0001-6050-7645]{David K. Sing}
\affil{Department of Earth \& Planetary Sciences, Johns Hopkins University, Baltimore, MD, USA}
\affil{Department of Physics \& Astronomy, Johns Hopkins University, Baltimore, MD, USA}

\author[0000-0002-3610-6953]{Jiayin Dong}
\altaffiliation{Flatiron Research Fellow}
\affil{Center for Computational Astrophysics, Flatiron Institute, New York, NY, USA}

\author[0000-0002-9328-5652]{Daniel Foreman-Mackey}
\affil{Center for Computational Astrophysics, Flatiron Institute, New York, NY, USA}

\author[0000-0003-3759-9080]{Tiffany Kataria}
\affil{Jet Propulsion Laboratory, California Institute of Technology, 4800 Oak Grove Drive, Pasadena, CA 91001}

\author[0000-0003-3726-5419]{Joanna K. Barstow}
\affiliation{School of Physical Sciences, The Open University, Walton Hall, Milton Keynes, MK7 6AA, UK}

\author[0000-0002-8515-7204]{Jayesh M. Goyal}
\affiliation{School of Earth and Planetary Sciences (SEPS), National Institute of Science Education and Research (NISER), HBNI, Jatni, Odisha 752050, India}

\author[0000-0002-8507-1304]{Nikole K. Lewis}
\affiliation{Department of Astronomy and Carl Sagan Institute, Cornell University, 122 Sciences Drive, Ithaca, NY 14853, USA}

\author[0000-0003-3667-8633]{Joshua D. Lothringer}
\affiliation{Department of Physics, Utah Valley University, Orem, UT, USA}

\author[0000-0001-6707-4563]{Nathan J. Mayne}
\affiliation{Department of Physics and Astronomy, Faculty of Environment Science and Economy, University of Exeter, EX4 4QL, UK}

\author[0000-0003-4328-3867]{Hannah R. Wakeford}
\affiliation{University of Bristol, HH Wills Physics Laboratory, Tyndall Avenue, Bristol, BS8 1TL, UK}

\author[0000-0002-4997-0847]{Duncan A. Christie}
\affil{Max Planck Institute for Astronomy, K\"{o}nigstuhl 17, D-69117 Heidelberg, Germany}

\author[0000-0003-4408-0463]{Zafar Rustamkulov}
\affil{Department of Earth \& Planetary Sciences, Johns Hopkins University, Baltimore, MD, USA}

\begin{abstract}

We present the first exoplanet phase curve measurement made with the \textit{JWST} NIRSpec instrument, highlighting the exceptional stability of this newly commissioned observatory for exoplanet climate studies. The target, WASP-121b, is an ultrahot Jupiter with an orbital period of 30.6\,hr. We analyze two broadband light curves generated for the NRS1 and NRS2 detectors, covering wavelength ranges of 2.70-3.72$\,\um$ and 3.82-5.15$\,\um$, respectively. Both light curves exhibit minimal systematics, with approximately linear drifts in the baseline flux level of 30\,ppm/hr (NRS1) and 10\,ppm/hr (NRS2). Assuming a simple brightness map for the planet described by a low-order spherical harmonic dipole, our light curve fits suggest that the phase curve peaks coincide with orbital phases $3.36 \pm 0.11 ^\circ$ (NRS1) and $2.66 \pm 0.12 ^\circ$ (NRS2) prior to mid-eclipse. This is consistent with the strongest dayside emission emanating from eastward of the substellar point. We measure planet-to-star emission ratios of $3,924 \pm 7 \,$ppm (NRS1) and $4,924 \pm 9 \,$ppm (NRS2) for the dayside hemisphere, and $136 \pm 8 \,$ppm (NRS1) and $630 \pm 10 \,$ppm (NRS2) for the nightside hemisphere. The latter nightside emission ratios translate to planetary brightness temperatures of $926 \pm 12 \,$K (NRS1) and $1,122 \pm 10 \,$K (NRS2), which are low enough for a wide range of refractory condensates to form, including enstatite and forsterite. A nightside cloud deck may be blocking emission from deeper, hotter layers of the atmosphere, potentially helping to explain why cloud-free 3D general circulation model simulations systematically overpredict the nightside emission for WASP-121b.
\end{abstract}


\section{Introduction} \label{sec:intro}

It is crucial to explore the coupling between the dayside and nightside hemispheres of tidally locked planets if we are to develop a global understanding of their atmospheres. One of the most effective means of constraining the chemical, thermal, and dynamical properties of both hemispheres is to measure planetary emission spectra over the course of a full planetary orbit \citep[e.g.][]{2007Natur.447..183K,2012ApJ...747...82C,2014Sci...346..838S,2018AJ....156...17K,2019A&A...625A.136A,2020AJ....160..137F,2020MNRAS.493..106I,2022NatAs...6..471M}. To this end, our team observed a spectroscopic phase curve for the ultrahot Jupiter WASP-121b using \textit{JWST} NIRSpec with the G395H grating. In this Letter, we present an analysis of two broadband light curves generated from this dataset, one for each of the NRS1 and NRS2 detectors. As well as providing some preliminary scientific results, our purpose is to provide a brief report on the stability of \textit{JWST} and NIRSpec when performing long-stare time-series measurements lasting tens of hours.

The target, WASP-121b, is an inflated gas giant with a mass of $1.183_{-0.062}^{+0.064} \, M_J$, a radius of $1.753 \pm 0.036\,R_J$ measured at red-optical wavelengths, and an orbital period of $30.59820 \pm 0.00001 \,$hr \citep{2016MNRAS.458.4025D, 2020A&A...635A.205B}. Secondary eclipse measurements made with \textit{HST} have shown that the dayside atmosphere has a thermal inversion, with a near-infrared brightness temperature close to 2,700\,K \citep{2017Natur.548...58E,2020MNRAS.496.1638M}. Evidence has been uncovered for numerous UV-optical absorbers in the atmosphere that are likely responsible for the dayside thermal inversion, such as VO, SiO, V, Fe, Mg, and Ca \citep[e.g.][]{2018AJ....156..283E,2020A&A...641A.123H,2021ApJ...914...12L}.

Phase curve measurements with \textit{HST} have shown that temperatures on the nightside of WASP-121b are likely low enough for refractory condensates containing V, Fe, Mg, and Ca to form \citep{2022NatAs...6..471M}. Since these latter elements are observed in the gas phase at the day-night terminator, if such clouds do form, they are presumably recirculated back to the dayside hemisphere where they are vaporized before settling to the deeper layers of the atmosphere. Meanwhile, nondetections of Ti and Al at the day-night terminator reported by \cite{2020A&A...641A.123H,2022arXiv221012847H} are consistent with these elements being cold-trapped in deeper layers of the nightside hemisphere as perovskite (CaTiO$_3$) and corundum (Al$_2$O$_3$).

From these previous studies, we are already observing the important connection between the dayside and nightside hemispheres of WASP-121b. One of the ultimate goals of our \textit{JWST} program will be to provide significantly tighter constraints on the dayside and nightside properties. This should be possible thanks to the \textit{JWST} data having a much higher signal-to-noise and spectral resolution than the existing \textit{HST} data, as well as the more favorable planet-to-star emission ratio at the longer wavelengths covered by NIRSpec.

\section{Observations and data reduction} \label{sec:obsdata}

\begin{figure}
\plotone{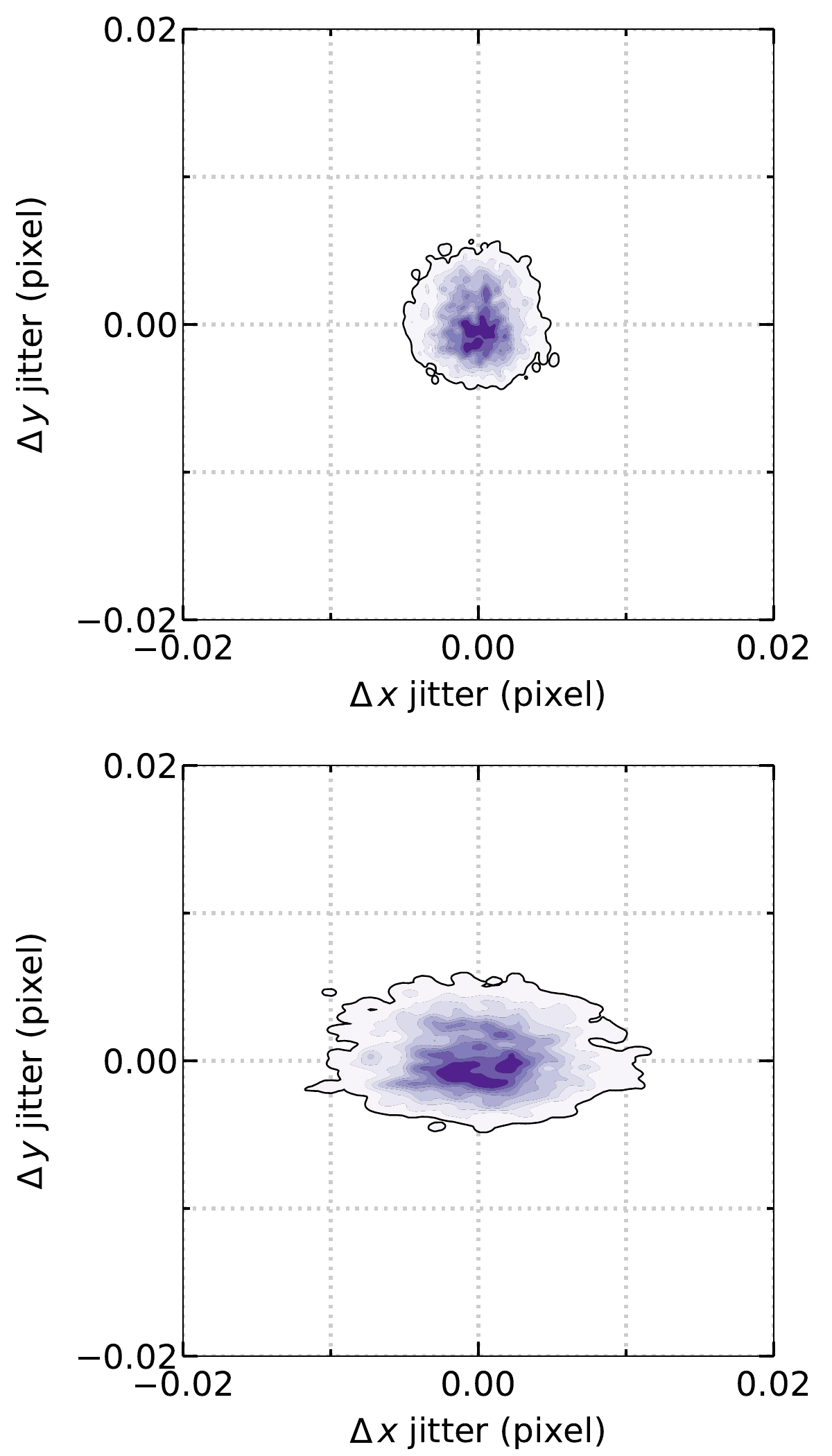}
\caption{Pointing jitter along the $x$ and $y$ axes of the NRS1 (top panel) and NRS2 (bottom panel) detectors, as measured by the FIREFly code.  \label{fig:xy}}
\end{figure}

Our team observed the WASP-121 system over a full planetary orbit on 2022 October 14-15 using the NIRSpec instrument as part of program GO-1729 (P.I.\ Mikal-Evans, co-P.I.\ Kataria). Observations were made without interruption for 37.8\,hr, commencing 145\,min prior to secondary eclipse ingress and continuing until 105\,min after egress of the following secondary eclipse. We used the G395H grating, providing wavelength coverage between 2.70-5.15$\,\um$ at $R \sim 3000$. For reading the detector, we employed the SUB2048 subarray option with the NRSRAPID readout pattern. Each integration consisted of 42 groups, translating to integration times of 38.8\,s and an overall duty cycle of 99\%. 

To extract the target spectra from the data frames, we used the (Fast InfraRed Exoplanet Fitting for Lightcurves) \texttt{FIREFLy} code \citep{2022ApJ...928L...7R,2022arXiv221110487R}. The data reduction begins with a customized reduction of stage 1 from the JWST \texttt{Python} pipeline \citep{2022zndo...7038885B}. Detector ``$1/f$'' noise \citep{2022A&A...661A..83B,2022ApJ...928L...7R,2020AJ....160..231S} is removed at the group level, masking out the spectral trace pixels and removing the median value from each column of the remaining background pixels. Bad pixels and cosmic rays are then flagged as outliers, both in the time-series and spatially on the 2D images. Movement of the spectra along the $x$- and $y$-axes of the detector are measured using cross-correlation, exhibiting standard deviations at the level of 0.002 pixels in both directions for NRS1 and in the $y$-direction for NRS2, while a higher standard deviation of 0.005 pixels is measured in the $x$-direction for NRS2 (Figure \ref{fig:xy}). This level of jitter is similar to previously reported on-sky performance \citep{2022arXiv220705632R,2022arXiv221101459E}, though demonstrated here on a much longer timescale. The 2D data frames are then adjusted using interpolation to account for the small pointing shifts. We find that this alignment procedure can dampen $x$ and $y$ correlations in the subsequent spectrophotometry, as it effectively takes into account the redistribution of flux across adjacent pixels. However, residual $x$ and $y$ trends may still remain due to intrapixel sentivities \citep{2022A&A...661A..83B}. Next, a fourth-order polynomial was used to trace the center of the point-spread-function (PSF) along the detector. Counts were then summed within an aperture centered on the trace with a width of 4.9 pixels for NRS1 and 5.5 pixels for NRS2, linearly interpolating the counts in those pixels where only a fraction was contained within the aperture. The aperture widths were selected after trialing a number of values and selecting those that minimized the scatter of the in-eclipse photometric time-series. To obtain the wavelength calibration, we extrapolated \textit{JWST} pipeline data products across the detector edge pixels that did not have an assigned wavelength. 

\section{Light curve analysis} \label{sec:lcanalysis}

\begin{figure*}
\epsscale{1.05}
\plotone{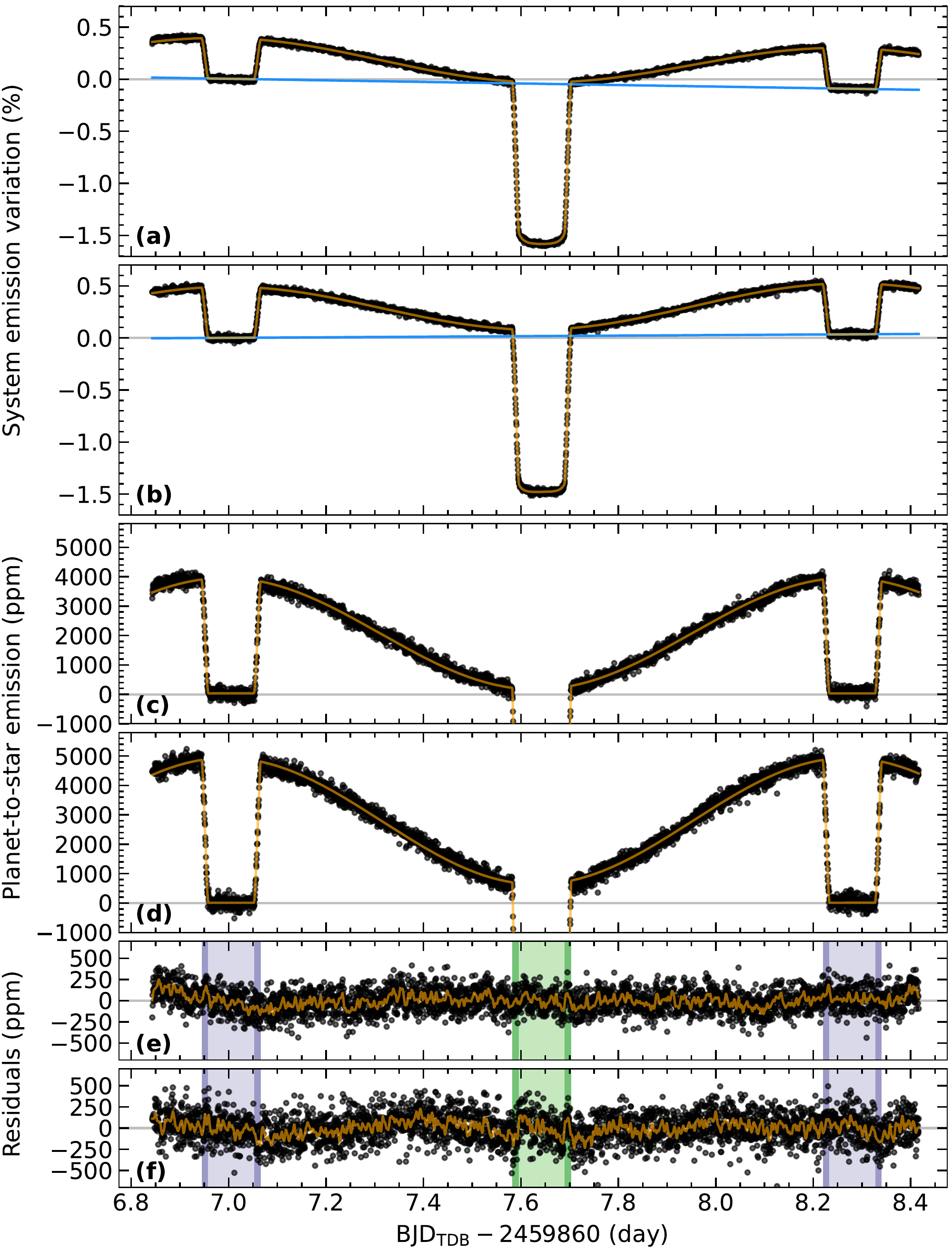}
\caption{\textit{Panels (a) and (b):} Raw broadband light curves for the NRS1 and NRS2 detectors, respectively. Orange lines show the best-fit light curve models. Blue lines show the corresponding instrument baseline trends. Gray horizontal lines are calibrated to the bottom of the first eclipse, to highlight the instrumental drift over the course of the observation. \textit{Panels (c) and (d):} The same light curves focusing on the planetary emission signal and after correcting for the instrument baseline trends shown by the blue lines in panels (a) and (b). \textit{Panels (e) and (f):} Black points show the residuals between the data and best-fit models for the NRS1 and NRS2 detectors, respectively. Orange lines show the residuals after applying a Gaussian filter to smooth the random noise. Green and purple shading show the times of transit and eclipse, respectively, with darker ranges corresponding to ingress and egress times. \label{fig:lc}}
\end{figure*}

We generated two light curves by summing each extracted 1D spectrum between dispersion columns 300-2042 for NRS1 and 5-2010 for NRS2, conservatively encompassing the full passbands of both detectors. The resulting raw light curves are shown in the top two panels of Figure \ref{fig:lc}. We modeled these light curves using the \texttt{starry} package \citep{2019AJ....157...64L} for the star-planet signal, multiplied by a linear trend in time to account for instrumental drift. For the planet brightness distribution, a simple dipole map was adopted, comprising the $Y_{0,0}$ and $Y_{1,0}$ spherical harmonics. For the stellar brightness distribution, a quadratic limb-darkening profile was assumed. In defining the data log-likelihood, we assumed that the data were normally distributed with standard deviations given by the pipeline Poisson uncertainties and an additional high-frequency systematics noise term combined in quadrature. 

We fitted the NRS1 and NRS2 light curves jointly, with the following parameters shared across both light curves: the stellar mass ($M_\star$) and logarithmic radius ($\log_{10} R_\star$); the differences in the orbital period ($\Delta P$) and transit mid-time ($\Delta T_c$) from the values reported by \cite{2020A&A...635A.205B}; the orbital inclination ($i$); and the planetary mass ($M_p$). Separate sets of the remaining model parameters were fitted for each light curve in the joint fit, namely: the stellar limb-darkening coefficients ($u_1$, $u_2$); the logarithmic planetary radius ($\log_{10} R_p$); the overall amplitude of the planetary brightness map ($A$) and the coefficient of the $Y_{1,0}$ spherical harmonic ($y_{1,0}$); the rotational offset of the planetary brightness map relative to a map with a brightness peak centered at the substellar point ($\Delta \phi$); the coefficients of the linear instrumental drift trend ($f_0$, $f_1$); and the high-frequency systematics noise term ($\sigma_{\rm{syst}}$). The priors we adopted for each parameter are shown by the dashed lines in Figure \ref{fig:pars}. For $M_\star$, $\log_{10} R_\star$, $i$, and $M_p$, these were normal priors based on the posteriors reported in \cite{2020A&A...635A.205B}. Broad normal priors were also adopted for $\log_{10} R_p$, $\Delta T_c$, $u_1$, $u_2$, $A$, $y_{1,0}$, $\Delta \phi$, $f_0$, and $f_1$. Half-normal priors only allowing positive values were adopted for $\sigma_{\rm{syst}}$. We marginalized the resulting posterior distribution using the No-U-Turn-Sampling \citep[NUTS;][]{nuts} method implemented by \texttt{PyMC3} \citep{pymc3}.

\section{Results} \label{sec:results}

\begin{figure*}
\plotone{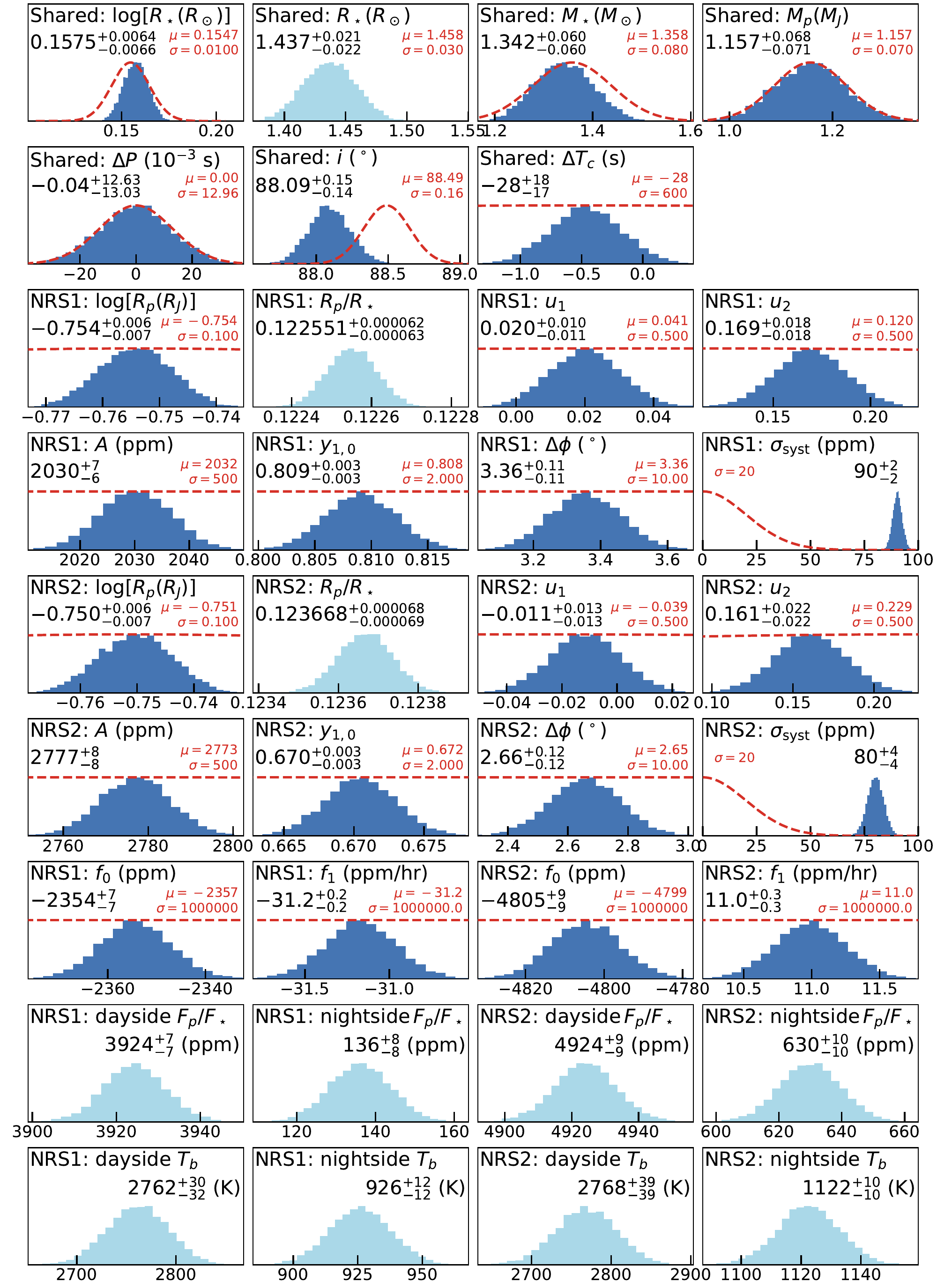}
\caption{Dark blue histograms show posterior distributions for model parameters that were varied in the light curve fitting. Light blue histograms show corresponding distributions for other parameters that were subsequently derived. Posterior medians and $\pm 34$\% credible intervals are listed in black font. Red dashed lines show the priors. For the normal priors, the mean $\mu$ and standard deviation $\sigma$ are listed in red font. Half-normal priors were adopted for the $\sigma_{\rm{syst}}$ parameters, allowing only positive values and with scale $\sigma$ listed in red font. Vertical axes are not labeled, as arbitrary normalizations have been applied to all plotted distributions. \label{fig:pars}}
\end{figure*}

The maximum a posteriori (MAP) light curve models are shown in Figure \ref{fig:lc}. Posterior distributions for each of the parameters that were varied in the fitting are shown as dark blue histograms in Figure \ref{fig:pars}, while the light blue histograms show the corresponding distributions for other parameters that were calculated from the latter, namely: the stellar radius $R_\star$; the planet-to-star radius ratio $R_p/R_\star$; the planet-to-star flux ratio $F_p/F_\star$ for the dayside and nightside hemispheres of the planet; and the corresponding planetary brightness temperatures $T_b$. The brightness temperatures were calculated by assuming $F_p/F_\star = (R_p/R_\star)^2\,B_p(T_b)/F_\star$, where $B_p$ denotes a blackbody function for the planetary flux, integrated over the instrument passband. In deriving distributions for $T_b$, uncertainties in the host star properties were accounted for by first randomly drawing values for the stellar effective temperature ($T_\star$), surface gravity ($\log g_\star$), and metallicity ([M/H]) from the posterior distributions reported in \cite{2016MNRAS.458.4025D}, which were assumed to be normal. For each of these sets of stellar properties, we used \texttt{pysynphot} \citep{2013ascl.soft03023S} to obtain a corresponding ATLAS9 stellar model \citep{2003IAUS..210P.A20C} that was then integrated over the instrument passband to obtain an estimate for $F_\star$. The resulting set of $F_\star$ values, along with the $R_p/R_\star$ and $F_p/F_\star$ posterior samples obtained from the light curve fits, were used to derive the final distributions for $T_b$.

We measure a significantly larger $R_p/R_\star$ value of $0.123668_{-0.000069}^{+0.000068}$ for NRS2, compared to $0.122551_{-0.000063}^{+0.000062}$ for NRS1. This difference can potentially be explained by the wavelength-dependent opacity of the planetary atmosphere. We find that the planetary brightness map is shifted eastward by $3.36 \pm 0.11 \, ^\circ$ ($30\sigma$) for NRS1 and by $2.66 \pm 0.12 \, ^\circ$ ($22\sigma$) for NRS2. We also make significant detections of the planetary nightside emission in both passbands: $136 \pm 8$\,ppm ($17\sigma$) for NRS1 and $630 \pm 10$\,ppm ($63\sigma$) for NRS2. These relative emission values translate to nightside brightness temperatures of $926_{-12}^{+12}\,$K and $1,122_{-10}^{+10}\,$K, respectively. For the dayside hemispheres, we obtain corresponding brightness temperatures of $2,762_{-32}^{+30}\,$K for the NRS1 passband and $2,768_{-39}^{+39}\,$K for the NRS2 passband. We note that the brightness temperature uncertainties are dominated by the stellar model uncertainties. When we only account for the uncertainties in $F_p/F_\star$ and $R_p/R_\star$ derived from the light curve fits while holding the stellar properties fixed to the best-fit values reported by \cite{2016MNRAS.458.4025D}, the uncertainties on the dayside brightness temperatures reduce by an order-of-magnitude from around 30-40\,K (Figure \ref{fig:pars}) to 3\,K. By contrast, the uncertainties associated with dayside brightness temperatures measured previously with the \textit{Spitzer Space Telescope} in the 3-5$\,\um$ wavelength range have been dominated by the $F_p/F_\star$ uncertainty. For example, \cite{2020AJ....159..137G} reported dayside brightness temperatures for WASP-121b of $2,490 \pm 77\,$K and $2,562 \pm 66\,$K in the $3.6\,\um$ and $4.5\,\um$ Infrared Array Camera (IRAC) passbands, which cover similar wavelength ranges to NRS1 and NRS2.

The long-term instrumental drift is found to be mild for both detectors. For NRS1, we obtain a negative drift of 1,172\,ppm over the course of the observation, corresponding to a rate of approximately 30\,ppm/hr. NRS2 is even better behaved, exhibiting a positive drift of only 416\,ppm from start to finish, at a rate of approximately 10\,ppm/hr. The standard deviations of the light curve residuals are 127\,ppm and 161\,ppm for the NRS1 and NRS2 detectors, respectively. This is 39\% and 17\% higher than the Poisson uncertainties derived from the instrument pipeline, respectively. As such, we derive high-frequency systematics noise values ($\sigma_{\rm{syst}}$) of $90 \pm 2$\,ppm for NRS1 and $80 \pm 4$\,ppm for NRS2.

\section{Discussion} \label{sec:discussion}

The small positive values that we measure for $\Delta \phi$ in both the NRS1 ($3.36 \pm 0.11 \, ^\circ$) and NRS2 ($2.66 \pm 0.12 \, ^\circ$) passbands translate to phase curves with maxima shifted prior to mid-eclipse. This in turn suggests that the eastern half of WASP-121b's dayside hemisphere is hotter on average than the western hemisphere, which is qualitatively consistent with the predictions of general circulation model (GCM) simulations of hot Jupiters \citep[e.g.][]{2002A&A...385..166S,2009ApJ...699..564S,2022ApJ...934...79K}. However, GCMs typically predict eastward phase curve shifts that are significantly higher than what we have measured here for WAP-121b; closer to $10^\circ$ or more \citep[e.g.][]{2018haex.bookE.116P}. Small phase curve shifts relative to GCM predictions have also been measured for a number of other hot Jupiters with properties similar to WASP-121b \citep[e.g.][]{2018AJ....155...83Z,2018AJ....156...17K}. For example, \cite{2018AJ....156...17K} measured an eastward phase shift of $2.0 \pm 0.7 \, ^\circ$ for WASP-103b in the \textit{Spitzer} IRAC $3.6\,\um$ channel, compared to a shift of $9.2^\circ$ predicted by a GCM presented in the same study. For ultrahot planets like WASP-121b and WASP-103b -- which both have dayside brightness temperatures well above $2,500\,$K -- Lorenz drag resulting from the motion of thermally ionized gas through the planetary magnetic field may inhibit the advection of heat throughout the atmosphere, reducing the observed phase curve offsets \citep{2010ApJ...719.1421P}. Alternatively, even if the hottest region of the dayside hemisphere is located significantly eastward of the substellar point, the presence of nightside clouds can reduce the observed phase shift in the resulting light curve \citep{2021MNRAS.501...78P}. We plan to investigate the latter possibility further in future analyses by including higher-order spherical harmonics terms for the planetary brightness map, which will allow for a sharper delineation in the brightness of the dayside and nightside hemispheres. 

\begin{figure}
\plotone{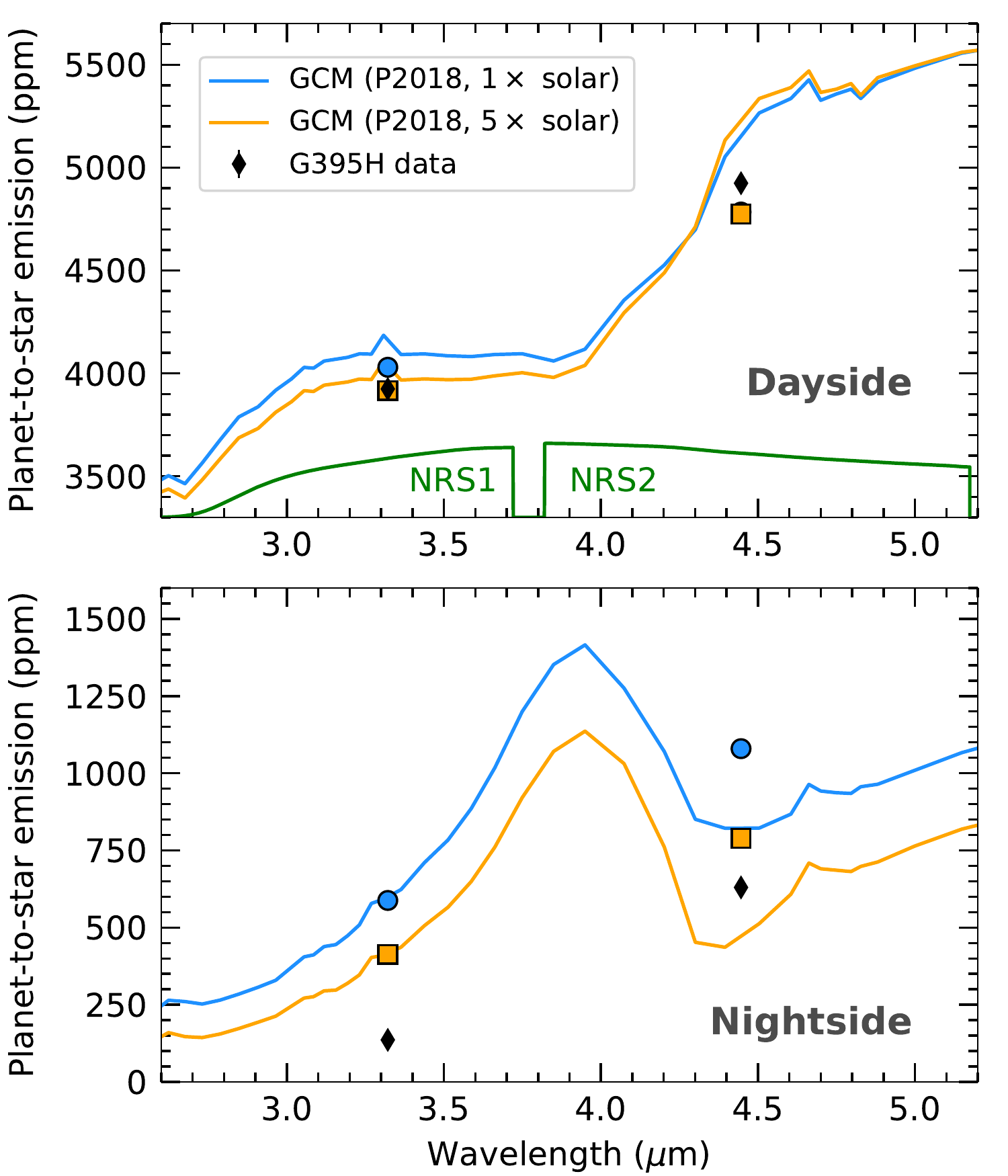}
\caption{Black diamonds show the measured dayside (top panel) and nightside (bottom panel) planet-to-star emission levels. \textit{Note that the measurement uncertainties are smaller than the diamond symbols.} Predictions from the cloud-free 3D GCM simulations of \citet{2018A&A...617A.110P} are also shown for heavy element enrichments of $1\times$ solar (blue lines) and $5\times$ solar (orange lines). Circle and square symbols show, respectively, the $1\times$ and $5\times$ solar GCM predictions binned to the light curve passbands, which are shown as green lines in the top panel. \label{fig:emission}}
\end{figure}

Our dayside and nightside emission measurements are plotted in Figure \ref{fig:emission} against the 3D GCM predictions for WASP-121b from \cite{2018A&A...617A.110P}, which assume $1\times$ and $5\times$ solar metallicity, chemical equilibrium, and do not include clouds. The models exhibit spectral features of H$_2$O and CO, which are in emission on the dayside due to a thermal inversion and absorption on the nightside due to a cooling temperature profile (Figure \ref{fig:emission}). The agreement between the measured dayside emission (3,924$\pm 7$\,ppm) and the $5\times$ solar GCM prediction (3,916\,ppm) across the NRS1 passband is very close to $1\sigma$. However, the dayside emission measured across the NRS2 passband (4,924$\pm 9$\,ppm) is $19\sigma$ higher than the $5\times$ solar GCM prediction (3,916\,ppm). For the nightside emission, the measurements across the NRS1 ($136 \pm 8 \,$ppm) and NRS2 ($630 \pm 10 \,$ppm) passbands are $>10\sigma$ lower than both the $1\times$ solar metallicity (588\,ppm for NRS1 and 1,080\,ppm for NRS2) and $5\times$ solar metallicity (413\,ppm for NRS1 and 789\,ppm for NRS2) GCM predictions. Such large discrepancies are unsurprising, given the small measurement uncertainties and the fact that the GCMs were not tuned to match the data. Still, it is notable that the measured nightside brightness temperatures are both substantially lower than predicted by the GCM simulations. Furthermore, the \cite{2018A&A...617A.110P} GCMs did not include the dissociation and recombination of hydrogen, which if anything would result in the predicted nightside temperatures being higher than those shown in Figure \ref{fig:emission} due to the release of latent heat \citep{2018ApJ...857L..20B,2019ApJ...886...26T,2022ApJ...934...79K}, exacerbating the discrepancy.

Clouds -- which were also not included in the \cite{2018A&A...617A.110P} GCMs -- might help explain our lower-than-predicted nightside brightness temperature measurements. Previous \textit{HST} observations have shown that the nightside temperature profile of WASP-121b does not have a thermal inversion at the near-infrared photosphere, instead cooling with decreasing pressure \citep{2022NatAs...6..471M}. If an optically thick cloud deck blankets the nightside hemisphere, it could block the emission from deeper, hotter layers of the atmosphere, lowering the observed brightness temperature \citep{2018AJ....155..150M,2019NatAs...3.1092K,2019AJ....158..166B,2021MNRAS.501...78P}. Even if the nightside cloud distribution is patchy rather than uniform, the brightness temperature can be reduced significantly relative to that of a cloud-free nightside \citep{2022ApJ...934...79K}.

The nightside brightness temperatures that we measure for WASP-121b in the NRS1 ($926_{-12}^{+12}\,$K) and NRS2 ($1,122_{-10}^{+10}\,$K) passbands fall below the condensation temperatures of silicates, such as enstatite and forsterite \citep{2010ApJ...716.1060V,2017MNRAS.464.4247W}, which are expected to be abundant in the nightside atmospheres of hot Jupiters \citep[e.g.][]{2021ApJ...918L...7G}. Therefore, our measurements reveal nightside conditions for WASP-121b that do appear to be conducive to the formation of silicate clouds, strengthening the possibility that clouds may prove key to understanding why our measured nightside brightness temperatures are significantly lower than predicted by the \cite{2018A&A...617A.110P} GCMs shown in Figure \ref{fig:emission}. Further work will be required to investigate if the difference in the nightside brightness temperatures obtained for the two passbands could be caused by the wavelength-dependent opacity of atmospheric layers above a cloud deck. For these future analyses, the dayside and nightside emission spectra will be extracted and interpreted, rather than the broad passbands considered here. 

It should also be stressed that the analysis we have presented here is preliminary. In particular, we have adopted a very simple light curve model and it is evident by eye that low-amplitude correlations still remain in the time-series of residuals (Figure \ref{fig:lc}). We did experiment with including the $x$ and $y$ pointing coordinates shown in Figure \ref{fig:xy} as additional linear decorrelation variables in the light curve fits, but found that this did not appreciably reduce the scatter in the residuals, nor did it affect the derived parameter distributions shown in Figure \ref{fig:pars}. Tidal deformation of WASP-121b could also subtly affect the observed phase curve \citep{2012ApJ...747...82C,2021ApJ...921..105W}. We did use \texttt{starry} to perform some preliminary light curve fits treating the planetary oblateness as a free parameter, but found that our basic conclusions were unaffected. Further work is required to determine if an oblate shape for the planet is supported by the data. Additional possibilities yet to be considered include adding higher-order spherical harmonics terms for the planetary brightness map and allowing for nonlinear instrumental baseline trends. These investigations are ongoing and could conceivably affect the constraints that we ultimately obtain for parameters of interest, such as the planetary nightside emission and brightness phase offsets.

In the meantime, our phase curve measurement for WASP-121b demonstrates the overall high level of stability that \textit{JWST} NIRSpec is capable of maintaining for a single-stare observation lasting 37.8\,hr. The approximately linear drift observed in the baseline flux level is extremely mild in comparison to the instrumental systematics that have affected past \textit{HST} and \textit{Spitzer} datasets \citep[e.g.][]{2012ApJ...747...82C,2014Sci...346..838S,2018AJ....156...17K,2022NatAs...6..471M}. We are hopeful that the additional 80-90\,ppm high-frequency systematics noise observed here in our preliminary analysis can be further reduced as more on-sky calibrations become available, along with continued refinement of the data analysis, such as improved treatment of the $1/f$ detector noise. Based on other recent analyses of NIRSpec data \citep{2022arXiv221110488A,2022arXiv221110487R}, we also anticipate that simple models fitted to light curves generated over narrower wavelength ranges than the broad passbands we have considered in the present study will achieve residual scatters closer to Poisson predictions.

\section{Conclusion} \label{sec:conclusion}

We have measured a full-orbit phase curve for the ultrahot Jupiter WASP-121b using \textit{JWST} NIRSpec. The resulting light curves generated across broad passbands for the NRS1 and NRS2 detectors exhibit minimal systematics over the 37.8\,hr observation. We find that the phase curve peaks are shifted prior to mid-eclipse by $3.36 \pm 0.11 \, ^\circ$ (NRS1) and $2.66 \pm 0.12 \, ^\circ$ (NRS2), suggesting that the eastern region of the dayside hemisphere is hotter on average than the western region. The measured dayside emission in the NRS1 passband is in good agreement with a cloud-free GCM assuming $5\times$ solar metallicity; however, the same GCM underpredicts the dayside emission in the NRS2 passband by $19\sigma$. For the nightside emission, cloud-free GCM simulations assuming $1\times$ and $5\times$ solar metallicity significantly overpredict the data. This observation could possibly be explained by nightside clouds blocking the emission from deeper, hotter layers of the atmosphere. The corresponding nightside brightness temperatures are $<1200$\,K in both passbands, which is cool enough for various condensates to form, including silicates such as enstatite and forsterite. \\


The authors are grateful to the anonymous referee for constructive feedback that improved the Letter. Support for JWST program GO-1729 was provided by NASA through a grant from the Space Telescope Science Institute, which is operated by the Association of Universities for Research in Astronomy, Inc., under NASA contract NAS 5-26555. J.K.B. was supported by a Science and Technology Facilities Council Ernest Rutherford Fellowship. N.M. was partly supported by a Science and Technology Facilities Council Consolidated Grant [ST/R000395/1], the Leverhulme Trust through a research project grant [RPG2020-82] and a UKRI Future Leaders Fellowship [grant number MR/T040866/1].

\software{ NumPy \citep{numpy2011}, SciPy \citep{2020SciPy-NMeth}, Matplotlib \citep{2007CSE.....9...90H}, JWST Python pipeline \citep{2022zndo...7038885B}, FIREFly \citep{2022ApJ...928L...7R}, starry \citep{2019AJ....157...64L}, PyMC3 \citep{pymc3}, pysynphot \citep{2013ascl.soft03023S} }

\facilities{ JWST(NIRSpec) }

All of the {\it JWST} data used in this paper were obtained from the Mikulski Archive for Space Telescopes (MAST) at the Space Telescope Science Institute. The specific observations analyzed can be accessed via \dataset[10.17909/23j6-ng29]{http://dx.doi.org/10.17909/23j6-ng29}

\bibliographystyle{aasjournal}
\bibliography{w121_jwst}

\end{document}